\documentclass[aps,preprint,nofootinbib]{revtex4}
\usepackage[T1]{fontenc}
\usepackage[utf8]{inputenc}
\usepackage{amsmath}
\usepackage{amssymb}
\usepackage{graphicx}
\usepackage{esint}
\usepackage{color}

\makeatletter
\@ifundefined{textcolor}{}
{%
 \definecolor{BLACK}{gray}{0}
 \definecolor{WHITE}{gray}{1}
 \definecolor{RED}{rgb}{1,0,0}
 \definecolor{GREEN}{rgb}{0,1,0}
 \definecolor{BLUE}{rgb}{0,0,1}
 \definecolor{CYAN}{cmyk}{1,0,0,0}
 \definecolor{MAGENTA}{cmyk}{0,1,0,0}
 \definecolor{YELLOW}{cmyk}{0,0,1,0}
}

\usepackage{amsfonts}
\usepackage{array}
\usepackage{multirow}
\usepackage{latexsym}
\usepackage{epsfig}
\usepackage{float}
\usepackage{dsfont}

\usepackage{ragged2e}
\justifying\let\raggedright\justifying
\usepackage{caption}
\providecommand{\tabularnewline}{\\}

\newcommand{\be}{\begin{equation}}
\newcommand{\ee}{\end{equation}}

\RequirePackage{doi}

\makeatother

\begin{document}

\title{Dark Energy and Neutrino Superfluids}

\author{Andrea Addazi}
\email{Addazi@scu.edu.cn}
\affiliation{Center for Theoretical Physics, College of Physics, Sichuan University,
Chengdu, 610064, PR China}
\affiliation{Laboratori Nazionali di Frascati INFN, Frascati (Rome), Italy}

\author{Salvatore Capozziello}
\email{capozzie@na.infn.it}
\affiliation{Dipartimento di Fisica ``E. Pancini'', 
Universit\`a di Napoli ``Federico II'' and 
Istituto Nazionale di Fisica Nucleare, Sezione di Napoli, Complesso Universitario di Monte S. Angelo, Via Cinthia, Ed. N I-80126 Napoli, Italy,}
\affiliation{Scuola Superiore Meridionale, Largo S. Marcellino 10, I-80138, Napoli, Italy,}

\author{Qingyu Gan}
\email{gqy@stu.scu.edu.cn}
\affiliation{Center for Theoretical Physics, College of Physics, Sichuan University,
Chengdu, 610064, PR China}
\affiliation{Scuola Superiore Meridionale, Largo S. Marcellino 10, I-80138, Napoli, Italy,}

\author{Antonino Marcian\`o}
\email{marciano@fudan.edu.cn}
\affiliation{Department of Physics \& Center for Field Theory and Particle Physics, Fudan University, 200433 Shanghai, China}
\affiliation{Laboratori Nazionali di Frascati INFN Via Enrico Fermi 54, Frascati (Roma), Italy}

\begin{abstract}
\noindent
We show that the neutrino mass, the dark matter and the dark energy can be explained in a unified framework, postulating a new invisible Born-Infeld field, which we name ``non-linear dark photon'', undergoing a meV-scale dynamical transmutation and coupled to neutrinos. Dark energy genesis is dynamically explained as a byproduct of the dark photon condensation, inducing the bare massless neutrinos to acquire an effective mass around the meV scale. It is fascinating to contemplate the channel induced by the non-linear dark photon leading to the pairing of the non-relativistic neutrinos, hence generating a cosmological superfluid state. As a consequence, the appearance of a light neutrino composite boson is predicted, providing a good cold dark matter candidate. In particular, if our model is enriched by an extra global Lepton number $U_{L}(1)$ symmetry, then the neutrino pair can be identified with a composite Majoron field with intriguing phenomenological implications for the neutrinoless-double-beta-decay $(0\nu\beta\beta)$. Our model carries interesting phenomenological implications since dark energy, dark matter and the neutrino mass are time-varying dynamical variables, as a consequence of the non-linear Born-Infeld interaction terms. Limits arising from \textbf{PLANCK+SNe+BAO} collaborations data are also discussed. Finally, our model allows for an inverse hierarchy of neutrino masses, with interesting implications for the \textbf{JUNO} experiment. 
\end{abstract}
\maketitle
\tableofcontents{}

\bigskip{}

\section{Introduction}

Neutrino mass, Dark Matter (DM) and Dark Energy (DE) are among the three most elusive as well as mysterious issues beyond the Standard Model (SM), yet, still at the center of physicists concerns and debates. We bravely and radically suspect that these three different aspects of Nature are intimately interconnected within a unified cosmological theory.

There have been many attempts to unify Particle Dark Matter and Neutrino mass (DM-Ne models) through symmetry extensions of the Standard model. For example, in the Majoron model, neutrino mass is generated due to the spontaneous symmetry breaking of the global Lepton number symmetry, while a new Nambu-Goldstone sterile particle provides a good candidate for Dark Matter \cite{Majoron1,Majoron2,Majoron3,Majoron4,berezinsky1993kev,barger1982majoron,berezhiani1992observable}. Moving from a complementary perspective, in the community of gravitation and cosmology there have been several searches for a common origin of Dark Matter and Dark Energy (DE-DM models) arising from extension of General Relativity (see Reviews  \citep{nojiri2007introduction,Nojiri:2010wj,Clifton:2011jh,Nojiri:2017ncd} and references therein).


In fact, there is very a attractive coincidence: the dark energy density is about $\rho_{DE}\sim\Lambda M_{Pl}^{2}\sim(1\,{\rm \mathrm{meV}})^{4}$, which is around the same neutrino mass scale that we expect for the lightest neutrino, i.e $m_{\nu}\sim1\,{\rm \mathrm{meV}}$.
  Is there any dynamical explanation for this intriguing accident in Nature? 
 On the other hand, a mystery of Nature puzzling  theoretical physicists remains: why neutrino physics and dark vacuum energy are so tiny compared to the natural scales of the Standard Model, such as the electroweak and the Quantum Chromodynamics energy scales? A unified picture for neutrino mass and dark energy genesis may indeed blow a fresh new wind over this urgently pressing problem, which has been widely investigated in many works \citep{fardon2004dark,gu2003dark,peccei2005neutrino,barger2005solar,brookfield2006cosmology,barbieri2005dark}.

\vspace{0.2cm}

In this paper, we show the existence of a generic minimal class of models for a DE-DM-Ne unification. We claim that in our model no new heavy hypothetical fermions, like Right-Handed (RH)  neutrinos, or extra Higgs bosons will be added to the Standard Model spectrum. Our assumptions will be only two, corresponding to just few free parameters:

$\,$

I) a novel fifth force spin-1 invisible interaction is introduced, with
the only additional desired requirement to be dynamically generating a condensate with an energy gap of $1\,{\rm \mathrm{meV}}$;

$\,$

II) neutrinos must be coupled with the new pseudo-vector
Born-Infeld (BI) boson.

$\,$

Within this minimal framework, we intuitively illustrate several important and unexpected consequences. First, it is quite immediate to realize that a new force, transmuting into a confining condensate with energy density of $1\,{\rm \mathrm{meV}}$, can provide a source for Universe acceleration (see \cite{addazi2016born} for a previous proposal following this way). On the other hand, even within the interplay of these two hypotheses, neutrinos can never propagate freely in the Universe: they rather interact with dark energy as an invisible electromagnetic-like background field. Therefore, neutrinos get a mass gap proportional to dark energy, as a byproduct of a frictional effect. In a certain sense, massless neutrinos acquire an effective mass, as much as electrons have a different effective mass in condensed matter
structures. 

Automatically, the neutrino mass happens to be tiny small as the
dark energy scale, without any Yukawa coupling fine-tuning or RH neutrino involved. It is worth to remark that instead of having fine-tuning on both dark energy and neutrino mass, here we need to assume one fine-tuning of the BI condensate. Neutrino mass is naturally tiny once the energy scale of the BI condensate is around meV scale. Thus fine-tuning in our model  is  on one rather than two parameters. One may argue that also the couplings among neutrinos and BI boson have to be fixed, but a $\mathcal{O}(1)$ coupling is considered as natural and closer to "familiar" Standard Model coupling constants.  Therefore, this is an alternative paradigm to the see-saw mechanism \citep{GellMann:1980vs,mohapatra1980neutrino,Sawada:1979dis}, in the ``healthy spirit" of Occam's razor logical principle, i.e it provides an economical explanation for DE and the neutrino mass generation. Then, we move on a final non-trivial step to the explanation of cold DM. We claim not to need any new extra sterile particle beyond the standard model in order to grasp the missing matter problem. In most of the astrophysical and laboratory cases, neutrino travels relativistically fast, close to the speed of light and, therefore, completely unbounded by the new invisible interaction. However, let us suppose to cool down neutrino to the meV scale: then neutrinos start being strongly coupled through the new vector boson and hence form Cooper pairs in a superfluid state. Neutrino superfluid is an old standing idea of ${\it Ginzburg}$ and ${\it Zhakarov}$ \citep{ginzburg1967superfluidity,ginzburg1969superfluidity}, which was later developed in many works \citep{caldi1999cosmological,kapusta2004neutrino,Dvali:2016uhn}. Nonetheless, we think that the power of this intuition was not fully appreciated by the community. If a copious amount of neutrinos was produced in the early Universe, then they could provide a sterile superfluid accounting for dark matter. We will provide qualitative arguments to support the hypothesis that dark matter is partially  (or possibly fully) composed of neutrino superfluid, and show that this can be easily accomplished in DE-DM-Ne framework. A quantitative estimate is possible but it is far beyond the main purposes of this paper. Indeed such estimation  is also dependent on the details of inflation reheating. Indeed, just as some axion models, one can also produce a neutrino superfluid from the decay of domain walls after inflation.

It is worth to mention that the current paper provides a much more accurate analysis and extensions of the  model proposed in Ref.\cite{addazi2016born}. 
	In particularly, we perform a quantitative analysis in comparison with data to show that a BI condensate can provide for dynamical dark energy from higher derivative terms of the BI action. Here, we reinterpret the neutrino condensate state as a composite Majoron, with intriguing implications for phenomenology in neutrinoless double beta decay searches. 
	This establishes an interesting connection among cosmological data and Universe acceleration with underground laboratory physics. Another important consequence of our model is that neutrinos necessarily have masses which are dynamically varying in time and this phenomenon could inspire future new experimental projects.

The plan of our paper is as follows: dark energy is discussed in section \ref{sec:Dark-Energy}; neutrino mass is dealt with in section \ref{sec:photon-Neutrino-Mass}; dark matter is investigated in section \ref{sec:Dark-Matter}; in Section V the relation between our model and the Majoron theory is studied; finally, in Section VI conclusions and remarks are offered\footnote{Throughout the paper, we use the Planckian natural units.}.

\section{Dark Energy}

\label{sec:Dark-Energy}

We formalise assumption (I) previously introduced by imposing the
condensation of the invisible gauge field, a dark photon, $\mathcal{A}_{\mu}$ in cosmological scale $M\simeq1\,\mathrm{meV}$, namely
\begin{equation}
({\rm I})\rightarrow\langle\mathcal{F}_{\mu\nu}\mathcal{F}^{\mu\nu}\rangle\sim M^{4},\label{first}
\end{equation}
where  {$\mathcal{F}_{\mu\nu}=\partial_\mu \mathcal{A}_\nu -\partial_\nu \mathcal{A}_\mu$  is the field strength of a new fifth force gauge boson field $\mathcal{A}_{\mu}$.
We postulate that the dark photon condensate provides the required contribution to dark energy. 
As in the case of QCD condensation, the non-vanishing value of $\langle \mathcal{F}^{2}\rangle$ emerges in the non-linear regime, where the perturbation theory breaks down. This provides a repulsive vacuum energy contribution, i.e. a candidate for DE  \citep{labun2010dark}. When the new vector field is simply abelian, the only possibility for the formation of the condensate is to resort to a non-linear higher derivative extension of the standard QED-like structure, e.g. as for the Born-Infeld theory.

The effects of a nonlinear electromagnetic theory in a cosmological setting have been studied by several authors (see e.g. Refs.~\citep{dona2015non,de2002nonlinear,labun2010dark,elizalde2003born,camara2004nonsingular,novello2007cosmological,novello2004nonlinear,kruglov2015universe,addazi2018dynamical}). In particular, it has been shown that a Born-Infeld field can provide a source for the Universe acceleration \citep{elizalde2003born}. In general a non-linear dark photon Lagrangian $\mathcal{L}_{\mathrm{eff}}$ coupled to gravity casts as follows
\begin{equation}
\mathcal{S}=\int\sqrt{-g}\left[-\frac{\mathcal{R}}{16\pi G}+\mathcal{L}_{\mathrm{eff}}[s,p]\right]d^{4}x,
\label{eq-Action}
\end{equation}
where
$\mathcal{L}_{\mathrm{eff}}[s,p]$ is a generic nonlinear effective
Lagrangian of the combination of dark gauge field strength $s=-\frac{1}{4}\mathcal{F}_{\mu\nu}\mathcal{F}^{\mu\nu}$ and
$p=-\frac{1}{4}\mathcal{F}_{\mu\nu}{}^{*}\mathcal{F}^{\mu\nu}=-\frac{1}{8}\mathcal{F}_{\mu\nu}\epsilon^{\mu\nu\rho\sigma}\mathcal{F}_{\rho\sigma}=-\frac{1}{8}\frac{1}{\sqrt{-g}}\mathcal{F}_{\mu\nu}\widetilde{\epsilon}^{\mu\nu\rho\sigma}\mathcal{F}_{\rho\sigma}$, where we also include the CP violating term in order to account for the more general unperturbed case. 

Since our main interest is the application to cosmology, we will work in the FLRW background. By definition, one can obtain the  stress-energy tensor $T_{\mu\nu }$ by varying action \ref{eq-Action} with respect to the metric $g^{\mu\nu }$, 
\begin{equation}
	T_{\mu \nu }  \equiv  \frac{-2}{\sqrt{-g}} \frac{\delta\left(\sqrt{-g} \mathcal{L}_{\mathrm{eff}}\right)}{\delta g^{\mu \nu}} 	  =  g_{\mu \nu} \mathcal{L}_{\text {eff }}-2 \frac{\partial \mathcal{L}_{\text {eff }}}{\partial s} \frac{\delta s}{\delta g^{\mu \nu}}-2 \frac{\partial \mathcal{L}_{\mathrm{eff}}}{\partial p} \frac{\delta p}{\delta g^{\mu \nu}} \nonumber. 
\end{equation}
From the stress-energy tensor $T_{\nu}^{\mu}$, and working within a frame that is co-moving with the fluid $\rho=-T_{0}^{0},P=T_{i}^{i}$, we have
\begin{eqnarray}
\rho & = & -\mathcal{L}_{\mathrm{eff}}-\mathcal{L}_{\mathrm{eff}}^{(1)}[s,p]\left(\mathcal{F}^{0\sigma}\mathcal{F}_{0\sigma}\right)+p\mathcal{L}_{\mathrm{eff}}^{(2)}[s,p],\\
P & = & \mathcal{L}_{\mathrm{eff}}+\mathcal{L}_{\mathrm{eff}}^{(1)}[s,p]\left(\mathcal{F}^{i\sigma}\mathcal{F}_{i\sigma}\right)-p\mathcal{L}_{\mathrm{eff}}^{(2)}[s,p],
\end{eqnarray}
where $\mathcal{L}_{\mathrm{eff}}^{(1)}[s,p]=\partial\mathcal{L}_{\mathrm{eff}}/\partial s$,
$\mathcal{L}_{\mathrm{eff}}^{(2)}[s,p]=\partial\mathcal{L}_{\mathrm{eff}}/\partial p$, 
and the index $i$ is not summed over.

The effective Lagrangian for the non-linear photon can include classical terms and quantum terms, which may provide the classical and quantum contributions to the condensation \citep{elizalde2003born}.
Due to the non-linear quantum effects, a condensation phenomenon found, namely
\begin{equation}
\langle\mathcal{F}_{\mu\nu}\mathcal{F}^{\mu\nu}\rangle_{Q}=\langle\mathcal{F}_{\mu\nu}{}^{*}\mathcal{F}^{\mu\nu}\rangle_{Q}=\alpha(t)\sim M(t)^{4},
\label{eq:quantum correlation}
\end{equation}
whose origins are purely of quantum nature. Within the FLRW spacetime, the condensate has the two components $\langle\mathcal{F}_{0\nu}\mathcal{F}^{0\nu}\rangle_{Q}=\alpha(t)/4$
and $\langle\mathcal{F}_{i\nu}\mathcal{F}^{j\nu}\rangle_{Q}=\delta_{i}^{j}\alpha(t)/4$. As an effect of the extra higher derivative terms, the cosmological constant will slowly run in time. Such a possibility  is not ruled out by  cosmic observations, such as  Supernovae Ia (SNe Ia) \citep{perlmutter1999measurements,riess1998observational}, cosmic microwave background (CMB) radiation \citep{spergel2003first,hinshaw2013nine}, large scale structure (LSS) \citep{tegmark2004cosmological,seljak2005cosmological}, baryon acoustic oscillations (BAO) \citep{eisenstein2005detection}
and weak lensing \citep{jain2003cross}, hence various  dynamical DE scenarios are proposed in literature  
\citep{zhao2017dynamical,alam2004case,clarkson2007dynamical,dent2011f,upadhye2005dynamical,xia2008constraints,guberina2006dynamical,farooq2017hubble,mainini2003modeling,sola2006dynamical,di2017constraining,bamba2012dark,addazi2018dynamical}.
On the other hand, the non-linear photon Lagrangian is in general not scale invariant at the classical level. Thus, for homogeneous and isotropic FLRW spacetimes, due to the equipartition principle electric and magnetic condensates acquire a stochastic background $\langle E_{i}E_{j}\rangle_{C}=\langle B_{i}B_{j}\rangle_{C}=\frac{1}{3}\epsilon(t)g_{ij}$, where $\epsilon$ is the classical radiation energy density \citep{tolman1930temperature,de2010nonsingular}. 
Indeed, the BI condensate receives two contributions: 1) the quantum correlator, which is purely quantum-mechanical in nature and that arises from the vacuum fluctuations; 2) the classical correlator, which accounts for the classical thermodynamics of the radiation. For the classical contribution, due to the isotropy of the spatial sections of FLRW geometry, one can perform the average procedure  as suggested by equipartition principle, where $C$ denotes an average over a volume that is relatively large compared to the wavelength while relatively small with respect to the curvature radius. 
Generically, we have that $E_{k}B^{k}=0$ implying $\langle E_{i}B_{j}\rangle_{C} $ vanishes. In terms of $\mathcal{F}_{\mu\nu}$, we have $\langle\mathcal{F}^{0\rho}\mathcal{F}_{0\rho}\rangle_{C}=-\epsilon(t),\,\langle\mathcal{F}^{i\rho}\mathcal{F}_{j\rho}\rangle_{C}=\frac{1}{3}\epsilon(t)\delta_{j}^{i}$
and $\langle \mathcal{F}_{\mu\nu}{}^{*}\mathcal{F}^{\mu\nu}\rangle_{C}=0$. Combining the quantum
and classical effects, we obtain
\begin{eqnarray}
\langle \mathcal{F}^{0\rho}\mathcal{F}_{0\rho}\rangle & = & \frac{\alpha(t)}{4}-\epsilon(t),\\
\langle \mathcal{F}^{i\rho}\mathcal{F}_{j\rho}\rangle & = & \left(\frac{\alpha(t)}{4}+\frac{\epsilon(t)}{3}\right)\delta_{j}^{i},\\
\langle \mathcal{F}_{\mu\nu}{}^{*}\mathcal{F}^{\mu\nu}\rangle & = & \alpha(t).
\end{eqnarray}
Thus, the fluid state parameter $w\equiv\langle P\rangle /\langle \rho \rangle$ for dark photon is found to be
\begin{equation}
w_{DE}(t)=\left.-1+\frac{16\epsilon(t)\mathcal{L}_{\mathrm{eff}}^{(1)}[s,p]}{-12\mathcal{L}_{\mathrm{eff}}[s,p]-3\left(\alpha(t)-4\epsilon(t)\right)\mathcal{L}_{\mathrm{eff}}^{(1)}[s,p]+12p\mathcal{L}_{\mathrm{eff}}^{(2)}[s,p]}\right|_{(s,p)=-\alpha(t)/4}.
\label{eq:state-parameter}
\end{equation}
Let us note that  $\mathcal{L}_{eff}/ \mathcal{L}_{eff}^{(1)}$ and $p\mathcal{L}_{eff}^{(2)}/ \mathcal{L}_{eff}^{(1)}$ are in the same order of $s$ and $p$, respectively. In general, for classical radiation dominating over the condensation,
i.e. $\epsilon>\!\!>\alpha$, we have $w\sim1/3$ as expected, while within the quantum effect dominant regime, where $\epsilon\rightarrow0$, we obtain $w\sim-1$. 
Beyond the quantum and classical limits, careful analysis must be devoted to the regimes corresponding to a quintessence-like condensate, characterised by an equation of state parameter $w>-1$, and to the phantom-like condensate, with $\omega<-1$, that appear in concrete realization of models described by a  specific form \footnote{It is worth to note that the entropy condition arsing from Holographic Naturalness highly restricts the possibility on various DE model, which shows that phantom cosmology is disfavored \citep{Addazi:2020vhq}.} of $\mathcal{L}_{\mathrm{eff}}$. Cosmological data do not rule out the possibility of $w<-1$, which will be also included for completeness in this work.

Let us consider now the specific case of the Born-Infeld theory, a no-ghost nonlinear electrodynamics, given by the effective action
\begin{equation}
\mathcal{L}_{\mathrm{BI}}[s,p]=\lambda\left[1-\sqrt{1-\frac{2s}{\lambda}-\frac{p^{2}}{\lambda^{2}}}\right].
\end{equation}
In practice, we can naturally set the dimensional coupling $\lambda=1\, \mathrm{meV}^{4}$, which is the scale of the dark energy density. By Eq.~(\ref{eq:state-parameter}), we obtain the equation of state
\begin{equation}
w(z)=-1+\frac{16\epsilon(z)}{3\left(4+\alpha(z)-\sqrt{16+8\alpha(z)-\alpha(z)^{2}}+4\epsilon(z)\right)},\label{eq:BI-w}
\end{equation}
where $z$ is the red-shift parameter. The evolution of the dark radiation $\epsilon(z)$ can be derived from the linear response of the fluctuation. One of the equation of motions of the BI action on the FLRW background is given by
\begin{equation}
\partial_{\mu}\left(a^{3}\frac{\mathcal{F}^{\mu\nu}+p{}^{*}\mathcal{F}^{\mu\nu}}{\sqrt{1-2s-p^{2}}}\right) =  0,\label{eq:BI-eom}
\end{equation}
which has a trivial solution $\mathcal{A}_{\mu}^{(0)}=0$. We consider the fluctuation around $\mathcal{A}_{\mu}^{(0)}$ as $\mathcal{A}_{\mu}=\delta \mathcal{A}_{\mu}+\left(\delta \mathcal{A}_{\mu}\right)^{2}+\cdots$, and keep it to the leading order in $\delta \mathcal{A}_{\mu}$. Then Eq.~(\ref{eq:BI-eom}) becomes $ \partial_{\mu}\left(a^{3}\mathcal{F}^{\mu\nu}/\lambda^{2}\right)=0 $.
As a simplifying ansatz, we can consider only time-dependent $\delta \mathcal{A}_{\mu}(t)$, and find the solution $\mathcal{F}_{0i}  =  \partial_{t}\delta \mathcal{A}_{i}=c/a$, where $c$ is a constant. With respect to the conformal time $\eta$, the result $\delta \mathcal{A}_{i}\sim c\eta$ indicates a linear perturbation. The radiation energy density is then found to be
\begin{equation}
\epsilon=\langle E_{i}E^{i}\rangle _{C}  =  \Big\langle \frac{1}{a^{2}}\mathcal{F}_{i0}\mathcal{F}_{i0}\Big\rangle_{C}=c_{0}+\frac{c^{2}}{a^{4}}\, , \label{eq:darkradiation}
\end{equation}
where the constant $c_{0}$ is introduced referring to  the  quantum radiative correction.

Let us first discuss the general physical picture of how $w$ evolves. The dark photon condensate appears when the temperature of the dark side decreases down to $1\, \mathrm{meV}$. We illustrate this toy model in Fig.~\ref{fig:wz}. One can see that with the expansion of the Universe, the condensation effect takes over the radiation, yielding the dark energy before the recombination.

\begin{figure}
\includegraphics[scale=0.85]{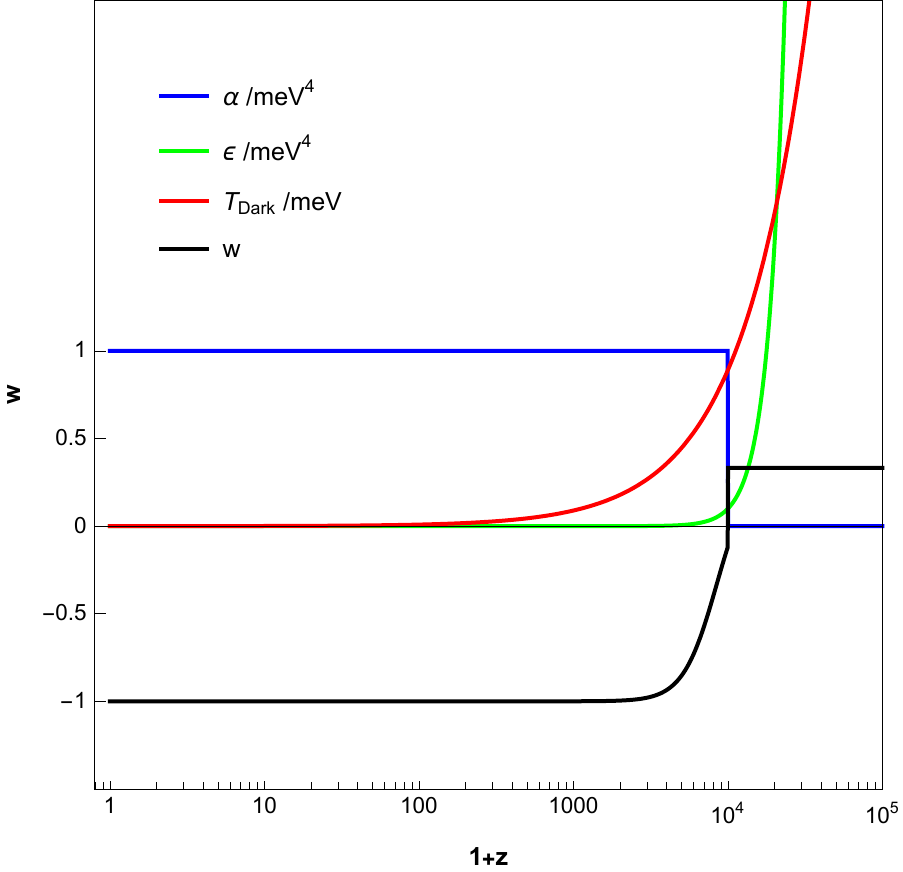}\,\,\,\,\,\,\, \includegraphics[scale=0.85]{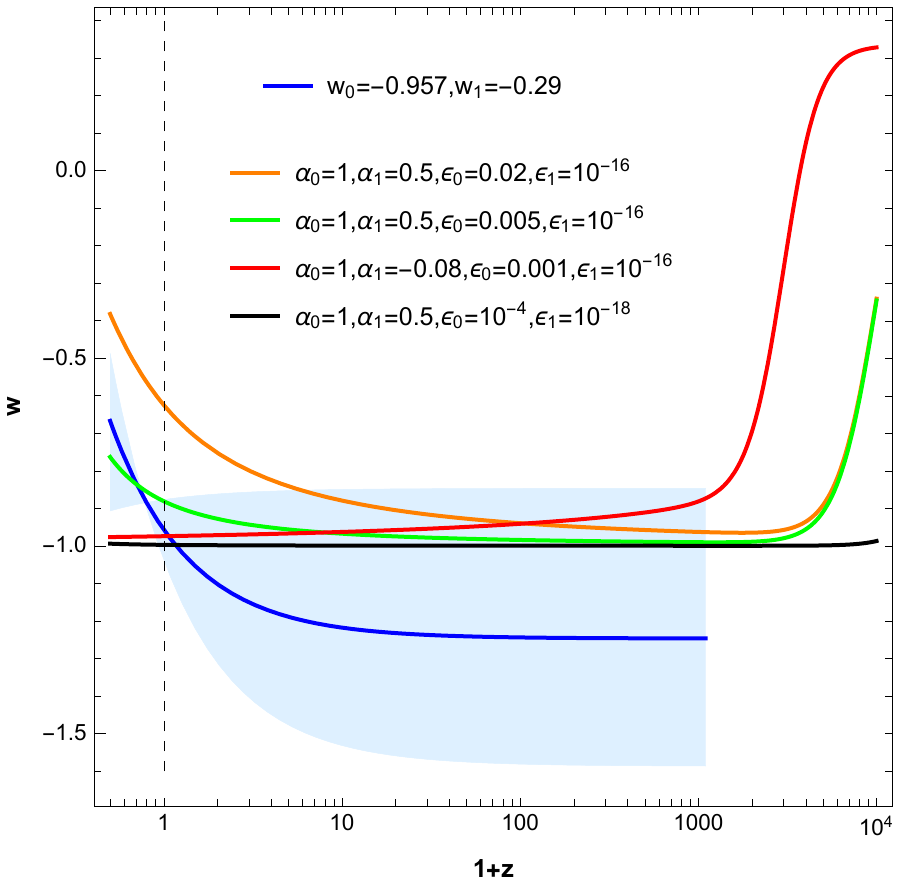}

\caption{$\mathbf{Left:}$  We show the evolution of the $w$ parameter as a function of the redshift variable $z$. The dark condensate dynamically transits from the radiation state ($w=1/3$) to the cosmological constant like state ($w=-1$). The condensate scale is set to a time-independent constant $1\, \mathrm{meV}^{4}$ after condensation. $\mathbf{Right:}$ The dark energy equation of state $w_{DE}$ is allowed at $1\sigma$ ($68.3\%$ confidence) in the shaded region by combined constraints from current data, assuming the Chevallier–Polarski–Linder (CPL) dynamical dark energy state \citep{aghanim2020planck}. Several different values
of $\{\alpha_{0},\alpha_{1},\epsilon_{0},\epsilon_{1}\}$ are set,
where the dark photon radiation remnant $\epsilon_{0}$ is constrained to a relatively small bound $\lesssim10^{-3}\, \mathrm{meV}^{4}$. }

\label{fig:wz}
\end{figure}

Many works have been devoted  on various specific parametrised forms of $w(z)$, e.g. the well known Chevallier-Polarski-Linder ansatz $w(z)=w_0+w_1 z/(1+z)$ \citep{chevallier2001accelerating,huterer1999prospects,jassal2005wmap,efstathiou1999constraining,seljak2005cosmological,upadhye2005dynamical}. In our case, we focus on the epoch $z\sim(-0.5,10000)$ during which the dark photon condensation is assumed to have been formed and slowly vary with time according to the ansatz $\alpha(z)=\alpha_{0}+\alpha_{1}\rm{log}(1+z)$, with $\alpha_{0} \sim\mathcal{O}(1)$ and $\alpha_{1} \sim\mathcal{O}(0.1)$, which is originated from one loop corrections to the dark photon propagator through a pair of neutrinos \cite{peskin}. 
Besides, from Eq. (\ref{eq:darkradiation}) the
dark radiation can be parametrised as $\epsilon(z)=\epsilon_{0}+\epsilon_{1}\left(1+z\right)^{4},(\epsilon_{0},\epsilon_{1}>0)$, where $\epsilon_{1}$ is constrained by $\lesssim10^{-16}$, since the dark energy is believed to exist in the CMB spectrum at $z\simeq1100$. In the right panel of the Fig.~\ref{fig:wz}, we explore some combinations of the parameter space $\{\alpha_{0},\alpha_{1},\epsilon_{0},\epsilon_{1}\}$ in comparison with the observational constraints arising from $Planck+SNe+BAO$ \cite{aghanim2020planck}.

From bayesian comparison with PLANCK data, we obtain as best fit the results in Table. \ref{tableI}.
We can see that our model has a $\Delta \chi^{2}$ which is comparable to $\Lambda CDM$.
This evidences that our model is a robust alternative to $\Lambda\, CDM$ model, while from the theoretical point of view a reduced fine-tuning in neutrino/BI scales as commented above. 

\begin{table}
\begin{tabular}{|c|c|c|c|c|c|c|c|}
\hline 
Likelihood & Frequency & Multipole range & $\chi^{2}$ & $\chi^{2}/N_{\textrm{dof}}$ & $N_{\textrm{dof}}$ & $\triangle\chi^{2}/\sqrt{2N_{\textrm{dof}}}$ & PTE{[}\%{]}\tabularnewline
\hline 
\hline 
\multirow{5}{*}{TT} & $100\times100$ & $30-1197$ & $1232.37$ & $1.06$ & $1168$ & $1.37$ & $8.66$\tabularnewline
\cline{2-8} \cline{3-8} \cline{4-8} \cline{5-8} \cline{6-8} \cline{7-8} \cline{8-8} 
 & $143\times143$ & $30-1996$ & $2032.45$ & $1.03$ & $1967$ & $1.08$ & $14.14$\tabularnewline
\cline{2-8} \cline{3-8} \cline{4-8} \cline{5-8} \cline{6-8} \cline{7-8} \cline{8-8} 
 & $143\times217$ & $30-2508$ & $2563.74$ & $1.04$ & $2479$ & $1.25$ & $10.73$\tabularnewline
\cline{2-8} \cline{3-8} \cline{4-8} \cline{5-8} \cline{6-8} \cline{7-8} \cline{8-8} 
 & $217\times217$ & $30-2508$ & $2549.66$ & $1.03$ & $2479$ & $1.00$ & $15.78$\tabularnewline
\cline{2-8} \cline{3-8} \cline{4-8} \cline{5-8} \cline{6-8} \cline{7-8} \cline{8-8} 
 & Combined & $30-2508$ & $2545.67$ & $1.03$ & $2479$ & $0.96$ & $16.81$\tabularnewline
\hline 
\multirow{7}{*}{TE} & $100\times100$ & $30-999$ & $1087.78$ & $1.12$ & $970$ & $2.70$ & $0.45$\tabularnewline
\cline{2-8} \cline{3-8} \cline{4-8} \cline{5-8} \cline{6-8} \cline{7-8} \cline{8-8} 
 & $100\times143$ & $30-999$ & $1031.84$ & $1.06$ & $970$ & $1.43$ & $7.90$\tabularnewline
\cline{2-8} \cline{3-8} \cline{4-8} \cline{5-8} \cline{6-8} \cline{7-8} \cline{8-8} 
 & $100\times217$ & $505-999$ & $526.56$ & $1.06$ & $495$ & $1.00$ & $15.78$\tabularnewline
\cline{2-8} \cline{3-8} \cline{4-8} \cline{5-8} \cline{6-8} \cline{7-8} \cline{8-8} 
 & $143\times143$ & $30-1996$ & $2027.43$ & $1.03$ & $1967$ & $0.98$ & $16.35$\tabularnewline
\cline{2-8} \cline{3-8} \cline{4-8} \cline{5-8} \cline{6-8} \cline{7-8} \cline{8-8} 
 & $143\times217$ & $505-1996$ & $1604.85$ & $1.08$ & $1492$ & $2.09$ & $2.01$\tabularnewline
\cline{2-8} \cline{3-8} \cline{4-8} \cline{5-8} \cline{6-8} \cline{7-8} \cline{8-8} 
 & $217\times217$ & $505-1996$ & $1430.52$ & $0.96$ & $1492$ & $-1.11$ & $86.66$\tabularnewline
\cline{2-8} \cline{3-8} \cline{4-8} \cline{5-8} \cline{6-8} \cline{7-8} \cline{8-8} 
 & Combined & $30-1996$ & $2045.11$ & $1.04$ & $1967$ & $1.26$ & $10.47$\tabularnewline
\hline 
\multirow{7}{*}{EE} & $100\times100$ & $30-999$ & $1026.79$ & $1.06$ & $970$ & $1.31$ & $9.61$\tabularnewline
\cline{2-8} \cline{3-8} \cline{4-8} \cline{5-8} \cline{6-8} \cline{7-8} \cline{8-8} 
 & $100\times143$ & $30-999$ & $1047.22$ & $1.08$ & $970$ & $1.78$ & $4.05$\tabularnewline
\cline{2-8} \cline{3-8} \cline{4-8} \cline{5-8} \cline{6-8} \cline{7-8} \cline{8-8} 
 & $100\times217$ & $505-999$ & $479.32$ & $0.97$ & $495$ & $-0.49$ & $68.06$\tabularnewline
\cline{2-8} \cline{3-8} \cline{4-8} \cline{5-8} \cline{6-8} \cline{7-8} \cline{8-8} 
 & $143\times143$ & $30-1996$ & $2001.70$ & $1.02$ & $1967$ & $0.54$ & $29.18$\tabularnewline
\cline{2-8} \cline{3-8} \cline{4-8} \cline{5-8} \cline{6-8} \cline{7-8} \cline{8-8} 
 & $143\times217$ & $505-1996$ & $1430.14$ & $0.96$ & $1492$ & $-1.11$ & $86.80$\tabularnewline
\cline{2-8} \cline{3-8} \cline{4-8} \cline{5-8} \cline{6-8} \cline{7-8} \cline{8-8} 
 & $217\times217$ & $505-1996$ & $1408.52$ & $0.94$ & $1492$ & $-1.51$ & $93.64$\tabularnewline
\cline{2-8} \cline{3-8} \cline{4-8} \cline{5-8} \cline{6-8} \cline{7-8} \cline{8-8} 
 & Combined & $30-1996$ & $1986.05$ & $1.01$ & $1967$ & $0.32$ & $37.16$\tabularnewline
\hline 
\end{tabular}

\caption{\it  The Goodness-of-fit tests using Planck temperature and polarization spectra, with the same methodology as $\Lambda CDM$ analysis performed by PLANCK collaboration.  $\Delta \chi^{2}=\chi^{2}-N_{\textrm{dof}}$ fitted to Planck TT+lowP with $N_{\textrm{eff}}$ is the number of degrees of freedom equal to the multipoles' number. The probability to exceed (PTE) the value of $\chi^{2}$ is in last column of table. A comparison with PLANCK analysis with $\Lambda CDM$ (see Refs.\cite{Planck:2015fie,Planck:2018vyg}) shows that our model has a comparable $\Delta \chi^{2}$ with data.}

\label{tableI}
\end{table}

Let us also mention that it is possible to introduce a kinetic mixing term between the non-linear dark photon and the
ordinary photon by means of the Pontryagin density $ \kappa  F^{\mu \nu} \mathcal{F}_{\mu \nu}$, where $F_{\mu \nu}$ is the strength tensor of the ordinary photon. Since our dark photon has an effective mass around the meV, the current limits on the kinetic mixing parameter arising from current photon mass limit is around $\kappa \lesssim 10^{-15}$ \cite{PDG}.

\section{Dark Photon Mass and Neutrino Mass}

\label{sec:photon-Neutrino-Mass}

The non-vanishing vacuum expectation of $\langle \mathcal{F}^{2}\rangle \sim M^{4}$ is related to the dark photon condensate with a mass gap $M\sim \mathrm{meV}$. We may then naturally identify the only energy scale entering the vacuum expectation, namely $M$, with the dark photon effective mass $m_{\mathrm{eff}}^{\gamma} \simeq M$. 

The vector vertex displayed in the left panel of Fig.~\ref{fig:effective-mass}, originating from the leading nonlinear term $\sim \mathcal{F}^{4}$ and with two legs representing the condensate background, could be interpreted as a dark photon propagating through the condensate ether. Consequently one can roughly read the Feynman diagram as $\sim M^{2} \mathcal{A}^{2}$, which shows that dark photon can obtain an effective mass around the meV scale. This effect may be understood in analogy to the gluon propagation in its condensate, with consequent acquisition of an effective mass gap related to the condensation scale.

Within the spirit of point (II), as stated in the Introduction, let us suppose that the Majorana neutrino $\nu$ couples to $\mathcal{A}_{\mu}$ as follows
\begin{equation}
({\rm II})\rightarrow\mathcal{L}_{int}=g\, \mathcal{A}_{\mu}\nu^{T}\mathcal{C}^{-1}\gamma^{5}\gamma^{\mu}\nu,
\label{eq:hypotheical 2}
\end{equation}
where $g$ is the coupling constant and $\mathcal{C}$ denotes the charge conjugate operation. One can see that the dark gauge potential $\mathcal{A}_{\mu}$ is a pseudo-vector. The interaction among propagating neutrinos and the background dark photon condensate medium would give to the bare massless neutrino an effective mass $m_{\mathrm{eff}}\simeq gM \simeq M$, where  $g$ is assumed to be a $\mathcal{O}(1)$ (see Fig.~\ref{fig:effective-mass}). Thus, in our model, neutrino mass and dark energy are interconnected  issues. 
We emphasize that $m_{\mathrm{eff}}$ is of the same order of magnitude as the time-dependent dark energy scale $M$. A large class of mass varying neutrino models was studied in Refs.~\citep{fardon2004dark,gu2003dark,peccei2005neutrino,barger2005solar,brookfield2006cosmology,barbieri2005dark}. Moreover, we mention that a scenario with Dirac neutrino rather than Majorana one was discussed in Ref.\cite{addazi2016born}.

\begin{figure}
\includegraphics[width=0.3\textwidth]{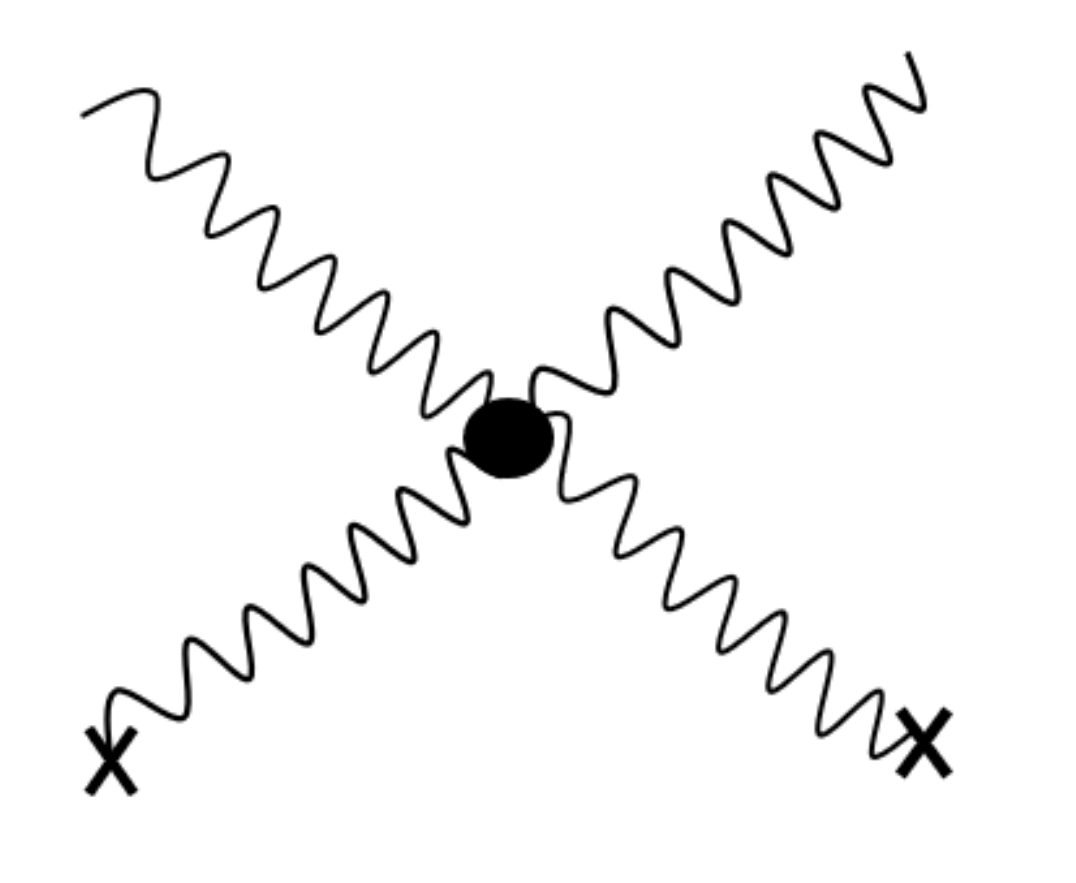}\,\,\,\,\,\,\includegraphics[width=0.4\textwidth]{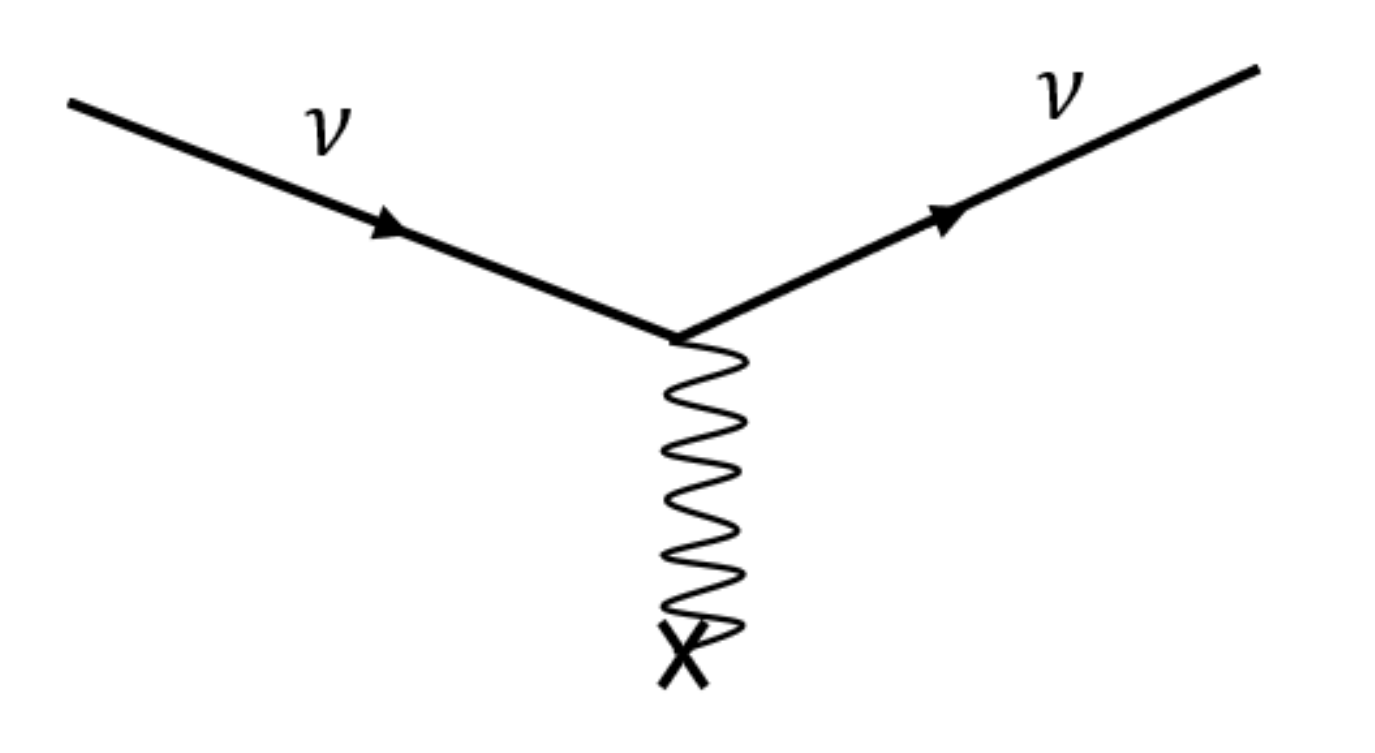}

\caption{{\bf Left: } Illustration of the generation of the dark photon's effective mass, resulting from to the interaction with the background dark photon condensate and mediated by the Euler-Heisenberg interaction term. The {\it cross} represents the background dark photon condensate. {\bf Right:} Illustration of the passage of neutrinos through the dark photon condensate, generating the neutrino's effective mass. }

\label{fig:effective-mass}
\end{figure}

Since neutrinos have different masses and oscillate among one another, it is quite natural to think that dark energy also generates neutrino mixings through neutrino flavour violating interactions. These latter are provided for instance by 
\begin{equation}
\mathcal{L}_{int}=g_{ff'}\mathcal{A}_{\mu}\nu_{f}^{T}\mathcal{C}^{-1}\gamma^{5}\gamma^{\mu}\nu_{f'},
\end{equation}
where $f$ and $f'$ are flavour indices and $g_{ff'}$ is a flavour
mixing matrix. The unitary transformation relating the flavour and
mass eigenstates is an analog of the Pontecorvo-Maki-Nakagawa-Sakata (PMNS) matrix \footnote{The idea that neutrino oscillations can be used to probe the dark energy was also explored in \citep{kaplan2004neutrino, Blasone:2004yh, Capolupo:2006et, Capolupo:2007hy, Capozziello:2013dja}.}.

In light of future neutrino experiments such as {\bf JUNO} \cite{An:2015jdp,Li:2014qca,Djurcic:2015vqa}, let us comment on the implications of our model. 
{\bf JUNO} is a middle-baseline antineutrino reactor, based on detection of antineutrinos generated by nuclear power sources.
Such a measure allows to determine the neutrino mass hierarchy with a promised significance of $4\sigma$ with six years of data taking. 
In particular, the high resolution measurement of the spectrum of antineutrinos 
can allow a precise determination of neutrino oscillations parameters, $\Delta m_{21}^{2}$, $\Delta m_{ee}^{2}$ and $\sin^{2} \theta_{12}$,
with $1\%$ precision. Such information is crucial in determining the sign of $\Delta m_{31}^{2}$:
if $m_{3}>m_{1}$ (normal hierarchy) or $m_{3}<m_{1}$ (inverse hierarchy). 
In our model, the neutrinos hierarchy follows the coupling matrix of neutrinos with the new interaction gauge bosons. 
Therefore, the model can allow for both inverse or normal hierarchies of neutrino masses. 
In our prospective, {\bf JUNO} would measure the hierarchy of interaction couplings of neutrinos with dark energy.
In particular, $m_{3}/m_{1}=g_{3}/g_{1}$, where $g_{1,3}$ are the coupling of first and third neutrinos with the BI field respectively. 
Thus an inverse hierarchy corresponds to the case $g_{1}>g_{3}$ while a normal one to $g_{3}>g_{1}$.


\section{Neutrino Condensation}

\label{sec:Dark-Matter}

In this section we specify the details of the non-relativistic neutrinos'  condensation, as triggered by the attractive force induced by the dark photon. In the UV regime, we consider the neutrino field to be a massless left-handed Weyl spinor $\xi_{\alpha}(x)$ embedded in the Majorana basis $\nu^{T}=\left(\begin{array}{cc} \xi_{\alpha}, & \xi^{\dagger\dot{\alpha}}\end{array}\right)$, coupling to $\mathcal{A}_{\mu}$ as Eq. (\ref{eq:hypotheical 2}), with the Lagrangian given  by
\begin{equation}
\mathcal{L}_{UV}=\frac{i}{2}\bar{\nu}\gamma^{\mu}\partial_{\mu}\nu+\frac{1}{2}g\mathcal{A}_{\mu}\bar{\nu}\gamma^{5}\gamma^{\mu}\nu.
\label{eq:UV Lag}
\end{equation}
The Lagrangian $\mathcal{L}_{UV}$ has a global $U(1)$ axial symmetry under the transformation  $\nu\rightarrow e^{i\theta \gamma^{5}}  \nu$.
A new effective 4-fermion interaction emerges once one integrates
out the vector field, leading to the non-vanishing expectation of
the Cooper pair $\langle\nu^{T}\mathcal{C}^{-1}\nu\rangle$, which
dynamically breaks the global $U(1)$ symmetry \footnote{Majorana neutrino condensation triggered by some scalar fields has been studied in \citep{bhatt2009majorana,antusch2003dynamical,barenboim2009inflation,fardon2004dark}.}.

\begin{figure}
\begin{centering}
\includegraphics{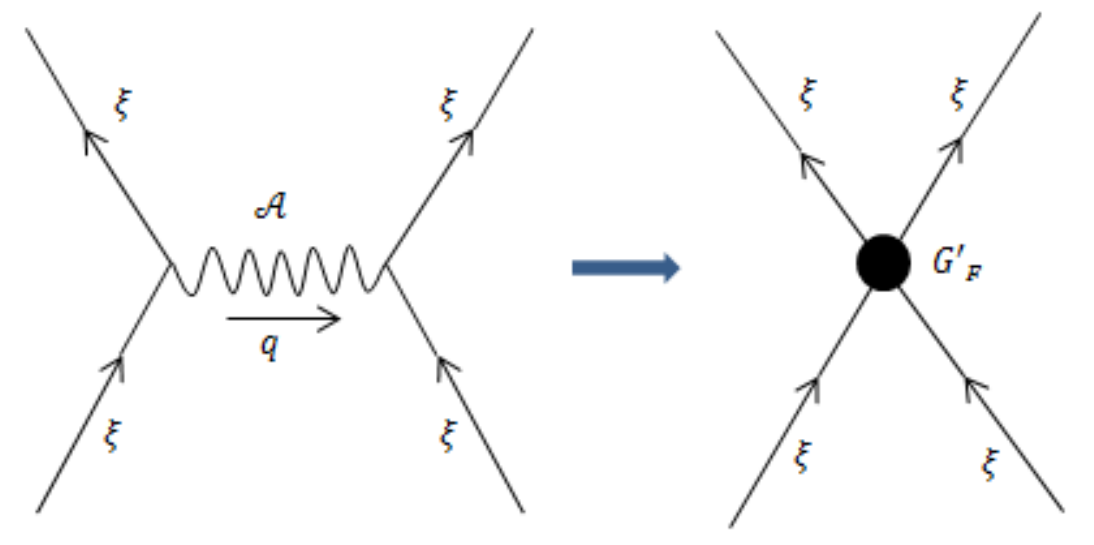}\includegraphics{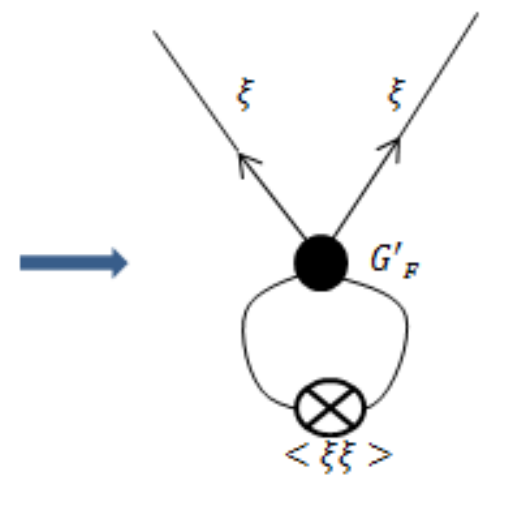}
\par\end{centering}
\begin{centering}
\caption{The left and middle diagrams illustrate the effective 4-fermion interaction of Majorana neutrinos induced by the invisible gauge field. The right side
diagram shows how the relativistic neutrinos (external
legs) obtain a mass term by propagating through the neutrinos
condensate (pairing legs).}
\label{fig:4fermion}
\par\end{centering}
\end{figure}

We now formalise the general picture discussed above. In terms of the two-component spinors, the Lagrangian Eq.~\eqref{eq:UV Lag} can be rewritten as
\begin{equation}
\mathcal{\mathcal{L}}_{UV}=i\xi^{\dagger}\bar{\sigma}^{\mu}\partial_{\mu}\xi+g\mathcal{A}_{\mu}\xi^{\dagger}\bar{\sigma}^{\mu}\xi \,.
\label{eq:interaction1}
\end{equation}
Here we take the van der Waerden notation \citep{dreiner2010two}.
As a local effective operator approach, integrating out the dark photon $\mathcal{A}_{\mu}$ we obtain a low energy effective 4-fermion interaction given by
\begin{equation}
\mathcal{L}_{int}\simeq\frac{g^{2}}{q^{2}+\left(m^{\gamma}_{\mathrm{eff}}\right)^{2}}\left(\xi^{\dagger}\bar{\sigma}^{\mu}\xi\right)\left(\xi^{\dagger}\bar{\sigma}_{\mu}\xi\right),
\label{eq:interaction2}
\end{equation}
with $m^{\gamma}_{\mathrm{eff}}\simeq \Lambda$ as argued in Sec.~\eqref{sec:photon-Neutrino-Mass}. For small transferred momentum  $q<\!\!<m^{\gamma}_{\mathrm{eff}}$, we obtain
\begin{equation}
\mathcal{L}_{int}  =  \frac{G'_{F}}{4}\left(\xi^{\dagger}\xi^{\dagger}\right)\left(\xi\xi\right),
\label{eq:interaction3}
\end{equation}
where  the Fierz identity $\bar{\sigma}^{\mu\dot{\alpha}\beta}\bar{\sigma}_{\mu}^{\dot{\gamma}\delta}=-2\epsilon^{\dot{\alpha}\dot{\gamma}}\epsilon^{\beta\delta}$
is utilized and we take the convention $\epsilon^{12}=-\epsilon^{21}=\epsilon_{21}=-\epsilon_{12}=1$.
Here we introduce the effective 4-ferimon interaction coupling $G'_{F}=8g^{2}/M^{2}$. The above process is illustrated
in Fig. \ref{fig:4fermion}.

Again, we stress that both non-relativistic and relativistic neutrinos acquire the effective mass $m_{\mathrm{eff}}\simeq M\simeq1\,{\rm \mathrm{meV}}$, but only non-relativistic neutrinos with kinetic energies $E<M$ undergo the condensation, while relativistic neutrinos with $E\gg M$ are practically unbounded from the condensate medium, and are mainly observed in astrophysical experiments. In the non-relativistic regime, we consider the fixed-axis spin states. Following \citep{dreiner2010two,anber2018new}, we introduce the non-relativistic spin-up and spin-down wave-functions $\psi_{\uparrow\downarrow}$, whose time derivatives are small compared to $m_{\mathrm{eff}}$. The non-relativistic scenario of the UV kinetic term can be casted in the Schr\"odinger field as $\underset{s=\uparrow\downarrow}{\sum}\psi_{s}^{*}\left(i\partial_{t}+\frac{\nabla^{2}}{2m_{\mathrm{eff}}}\right)\psi_{s}(x)$,
and  the UV interaction term Eq. (\ref{eq:interaction3}) reduces to $G'_{F}\psi_{\uparrow}^{*}\psi_{\downarrow}^{*}\psi_{\downarrow}\psi_{\uparrow}$.
Taking into account the chemical potential $\mu$, neutrinos
cooled enough by the dark photon background are described by the non-relativistic
action
\begin{equation}
\mathcal{S}[\psi^{\dagger},\psi]  =  \int d^{4}x\left[\underset{s=\uparrow\downarrow}{\sum}\psi_{s}^{*}\left(i\partial_{t}+\frac{\nabla^{2}}{2m_{\mathrm{eff}}}+\mu\right)\psi_{s}(x)+G'_{F}\psi_{\uparrow}^{*}\psi_{\downarrow}^{*}\psi_{\downarrow}\psi_{\uparrow}(x)\right]\, . \label{eq:action-cooper}
\end{equation}
This is reminiscent of the BCS theory, although here the attractive force regulated by the bare coupling constants $G'_{F}>0$, and triggering the generation of the condensate, is induced by the dark photon \footnote{A similar scenario for the vacuum energy condensation was studied in \citep{Dona:2016fip, Addazi:2017qus}, assuming torsional gravity in order to derive an attractive super-conducting behaviour.}.

Since the attractive channel is mediated by the dark photon,
let us proceed with the standard approach to model superfluid condensation \citep{altland2010condensed,casalbuoni2003lecture,fai2019quantum,popov1991functional,schmitt2015introduction}.
We focus on the Cooper pairs with zero spin fermion-fermion bilinear $\psi_{\uparrow}\psi_{\downarrow}$, which serves as the order parameter of the neutrino s-wave superfluid \citep{Schakel:1999pa}.
Applying the Hubbard-Stratonovich identity to an auxiliary complex scalar field $\Phi(x)$ in the functional integral $Z=\int\mathcal{D}\psi\mathcal{D}\psi^{*}e^{i\mathcal{S}}$, we obtain
\begin{eqnarray}
Z & = & \frac{1}{Z_{0}}\int\mathcal{D}\psi\mathcal{D}\psi^{*}\mathcal{D}\Phi\mathcal{D}\Phi^{*}\exp\left(i\mathcal{S}'\left[\psi,\psi^{*},\Phi,\Phi^{*}\right]\right), \label{eq:functional-HS} \\
\mathcal{S}'\left[\psi,\psi^{*},\Phi,\Phi^{*}\right] & = & \mathcal{S}_{0}\left[\psi,\psi^{*}\right]+\int d^{4}x\left[-\frac{|\Phi|^{2}}{G'_{F}}-\Phi\left(\psi_{\uparrow}^{*}\psi_{\downarrow}^{*}\right)-\Phi^{*}\left(\psi_{\downarrow}\psi_{\uparrow}\right)\right],
\label{eq:partition-HS}
\end{eqnarray}
where $Z$ is normalized to the free theory $Z_{0}=\int\mathcal{D}\psi\mathcal{D}\psi^{*}e^{i\mathcal{S}_{0}}$.
By varying   $\mathcal{S}'$ with respect to $\Phi$, we obtain the equation of motion
$\Phi=G'_{F}\psi_{\uparrow}\psi_{\downarrow}$, which indicates that the composite scalar  $\Phi$ can be interpreted as the neutrino pairing field. In terms of  the  Nambu-Gorkov
basis  $\chi^{T}=\left(\begin{array}{cc}
\psi_{\uparrow}, & \psi_{\downarrow}^{*}\end{array}\right)$, the action in Eq. (\ref{eq:partition-HS}) can be arranged into
\begin{equation}
\mathcal{S}'\left[\chi,\chi^{\dagger},\Phi,\Phi^{*}\right]=\int d^{4}x\left(\chi^{\dagger}\mathcal{M}^{-1}\chi-\frac{|\Phi|^{2}}{G'_{F}}\right),
\label{eq:action-HS}
\end{equation}
with
\begin{equation}
\mathcal{M}^{-1}=\left(\begin{array}{cc}
i\partial_{t}+\frac{\nabla^{2}}{2m_{\mathrm{eff}}}+\mu & -\Phi\\
-\Phi^{*} & i\partial_{t}-\frac{\nabla^{2}}{2m_{\mathrm{eff}}}-\mu
\end{array}\right)\label{eq:M-matrix}\, .
\end{equation}
Plugging Eq.~(\ref{eq:action-HS}) into the partition function Eq.~(\ref{eq:functional-HS}), and integrating over $\chi$ by means of the Gau\ss~integral, we obtain the one-loop effective action
\begin{eqnarray}
\frac{Z}{Z_{0}} & = & \intop\mathcal{D}\Phi\mathcal{D}\Phi^{*}e^{i\mathcal{S}_{\mathrm{eff}}\left[\Phi,\Phi^{*}\right]},
\label{eq:eff-partition}\\
\mathcal{S}_{\mathrm{eff}}\left[\Phi,\Phi^{*}\right] & = & -\frac{i}{2}\textrm{Tr}\left[\log\left(\mathcal{M}_{0}\mathcal{M}^{-1}\right)\right]-\int d^{4}x\frac{|\Phi|^{2}}{G'_{F}}.
\label{eq:eff-action}
\end{eqnarray}
Here $\mathcal{M}_{0}^{-1}$ corresponds to the free theory $Z_{0}=\left[\textrm{det}\:\mathcal{M}_{0}^{-1}\right]^{1/2}$, while the trace Tr$[...]$ is taken with respect to the Nambu-Gorkov space. For the superfluid ground state,  the paring field $\Phi(x)$ obtains a static and uniform gap $\Delta$, which is assumed to be real. Within the saddle-point approximation, $Z\propto e^{i\mathcal{S}_{\mathrm{sp}}}$ --- with  $\mathcal{S}_{\mathrm{sp}}[\Delta]=\mathcal{S}_{\mathrm{eff}}\left[\Delta,\Delta^{*}\right]$ --- can be treated as the thermodynamic grand potential. Therefore the
dependence of $\Delta$ on the thermodynamic variables can be extracted
from the extreme condition $\delta \mathcal{S}_{\mathrm{sp}}/\delta\Delta=0$,
which yields the gap equation
\begin{equation}
\Delta=\frac{G'_{F}}{4}\int\frac{d^{3}\mathbf{p}}{\left(2\pi\right)^{3}}\frac{\Delta}{\sqrt{\left(\frac{\mathbf{p}^{2}}{2m_{\mathrm{eff}}}-\mu\right)^{2}+\Delta^{2}}} \,.
\label{eq:gap-equation}
\end{equation}
A full non-perturbative and non-trivial solution of Eq.\ref{eq:gap-equation} can only be found with numerical methods.
 For the positive coupling $G'_{F}$, one can see that Eq.(\ref{eq:gap-equation}) will always possess some nonzero gap solution $\Delta >0$, which implies the formation of neutrino condensates (see for example Refs.\citep{altland2010condensed,casalbuoni2003lecture}).
A simple BCS-like estimate provides the size of the condensate, namely  \citep{caldi1999cosmological}
\begin{equation}
\Delta\sim p_{F}\exp\left(-\frac{1}{p_{F}^{2}G'_{F}}\right),
\end{equation}
where $p_{F}=\sqrt{2\mu m_{\mathrm{eff}}}$ is the Fermi momentum for the non-relativistic neutrinos. Let us note that the chemical potential $\mu$ can be interpreted as the finite density parameter of the condensate or the Fermi energy level, which is around the meV scale. Taking into account $\mu\simeq M$,
$m_{\mathrm{eff}}\simeq g M$ and $G'_{F}\simeq g^{2}/M^{2}$, we have $\Delta\sim\sqrt{g}M\exp\left(-1/g^{3}\right)\sim\ M$, provided that we assume $g\sim\mathcal{O}(1)$. Furthermore, the critical temperature is of the same order of the  gap scale, i.e. meV.

Let us comment on the dynamical generation of relativistic neutrino's effective mass. In Sec.~\eqref{sec:photon-Neutrino-Mass}, we have found that the relativistic neutrino propagating in the dark photon condensate would acquire a mass term. Thus the
ground state of the pair field $\langle\Phi\rangle=G'_{F}\langle\psi_{\uparrow}\psi_{\downarrow}\rangle=\Delta$
implies the non-vanishing expectation value of $\langle\xi\xi\rangle=\langle\xi^{\dagger}\xi^{\dagger}\rangle\simeq\Delta/G'_{F}$.
Consequently, in Fig.~\ref{fig:4fermion} the 4-fermion interaction that accounts for two legs pairing as a condensate, reads $G'_{F}\xi^{\dagger}\xi^{\dagger}\langle\xi\xi\rangle+G'_{F}\langle\xi^{\dagger}\xi^{\dagger}\rangle\xi\xi\simeq\Delta\left(\xi^{\dagger}\xi^{\dagger}+\xi\xi\right)\simeq\Delta\nu^{T}\mathcal{C}\nu$, which indicates the emergence of an effective Majorana neutrino mass $m_{\mathrm{eff}}\simeq\Delta\sim M$.

We may finally discuss the superfluid properties of the neutrino condensate. As it is well known, a superfluid originates from the spontaneous symmetry breaking of an abelian global $U(1)$. We observe that the global axial $U_{A}(1)$ symmetry of the action Eq.~(\ref{eq:action-cooper}) is broken by the formation of a ground state with fixed global phase, i.e. $\Delta\in\mathbb{R}$. The fluctuation around the ground state, namely $\Phi(x)=\left(\Delta+\delta\rho(x)\right)e^{i\delta\theta(x)}$, will produce two collective modes: the massive mode $\delta\rho$ and a massless mode $\delta\theta$. In the vicinity of the critical temperature, the dynamics of $\Phi(x)$ is described by the time-dependent Ginzburg-Landau Lagrangian
\begin{equation}
\mathcal{L}_{GL}[\Phi,\Phi^{*}]=i\Phi^{*}\partial_{t}\Phi-\frac{1}{2m_{\Phi}}\mathbf{\nabla}\Phi^{*}\mathbf{\nabla}\Phi-\alpha\Phi^{*}\Phi-\frac{1}{2}\beta\left(\Phi^{*}\Phi\right)^{2}\,.
\label{eq:GL}
\end{equation}
Here $m_{\Phi}$ is the effective mass of the composite neutrino pairs, which can be roughly parametrised as $m_{\Phi}\simeq2m_{\mathrm{eff}}$. A detailed calculation relates the coefficients $\alpha$, $\beta$ to the thermal parameters of the system \citep{casalbuoni2003lecture}. Below the critical temperature, the phase fluctuation $\delta\theta(x)$ is un-gapped and fulfils in the low-energy regime the linear dispersion
\begin{equation}
\omega_{\mathbf{k}}^{2}=-\frac{\alpha}{\beta}\frac{\mathbf{k}^{2}}{2m_{\Phi}}\sim\frac{\Delta}{m_{\mathrm{eff}}}\mathbf{k}^{2}\,.
\end{equation}
Therefore, un-gapped massless mode propagates as phonon like excitation, which may relate Dark matter to a superfluid state having an extra long-range interaction and effectively modifying the newtonian potential \cite{khoury2016dark,Addazi:2018ivg,Sharma:2018ydn,Ferreira:2018wup,Berezhiani:2018oxf,Famaey:2019baq}.

Neutrino superfluid has many interesting applications in cosmological and astrophysical phenomena, and include the formation of neutrino vortices, exotic neutrino superfluid boson stars, gravitational waves etc., which have been extensively investigated in a wide literature --- see e.g. Refs.~\citep{berezhiani2015theory,volovik2001superfluid,khoury2016dark,Addazi:2018ivg,Sharma:2018ydn,Ferreira:2018wup,Berezhiani:2018oxf,Famaey:2019baq}.

\section{Composite Majoron}


Let us now postulate an extension of the SM symmetry with an extra global $U_{L}(1)$. The BI field condensation induces the Majorana neutrino pairing into Cooper pairs. This phenomenon dynamically breaks the global lepton number, since the neutrino condensate carries a double charge unit with respect to the Lepton-symmetry. The dynamical symmetry breaking of the $U_{L}(1)$ generates the neutrino mass term. A composite pseudo-Nambu-Goldstone is then obtained in the model, and is related to the spontaneous symmetry breaking of the $U_{L}(1)$ symmetry. This particle is then identified with the Majoron field \citep{Majoron1,Majoron2,Majoron3,Majoron4}.

The condensation process induces a Majorana mass term, which is in turn related to the emergence of a Nambu-Jona-Lasinio(NJL) four fermion interaction during the BI condensation, namely 
\begin{equation}
\label{BIAI}
\mathcal{L}_{NJL,\nu}=G'_{F} \langle \nu\nu\rangle \nu\nu+h.c. \rightarrow \Delta L=2 \,.
\end{equation}

In the model we have been introducing, the natural mass scale of the Majoron is the meV scale. In this case, the Majoron field does not remain massless, and can only compose a superfluid state. In other words, the composite Majoron can provide a viable candidate of superfluid Dark Matter. Such a model seems compatible with the recent proposal of cold dark matter composed by light Majorons, as proposed in Ref.~\cite{Reig:2019sok}. Within this perspective, the right amount of Majoron Dark Matter can be generated from the decay of topological defects, produced in the dark sector and decaying into composite states \cite{Reig:2019sok}.

Our model poses further questions for the phenomenology of the neutrino-less double beta ($0\nu\beta\beta$) decay. As it is well known, in the usual Majorana mass models, the neutrino-antineutrino identification leads to reconnection of the neutrino-antineutrino internal line from two simultaneous beta decays. Within the case of the traditional fundamental Majoron, an invisible particle emission can alter the statistical distribution of the emitted electrons as a multi-body decay. But in our case, the Majoron would be emitted at energies around the nuclear KeV scale, where neutrinos are practically un-bounded \citep{barger1982majoron}. Thus, in our scenario, the Majoron emission would be substituted by the emission of a couple of electronic (anti)neutrinos. 

If this would be the case, we would arrive to the astonishing conclusion that the composite Majoron is completely invisible in the $0\nu\beta\beta$ decays. This prediction will be eventually relevant for the next generation of experiments on the $0\nu\beta\beta$ decay, including {\bf LEGEND}, {\bf CUORE}, {\bf nEXO}, {\bf GERDA-\uppercase\expandafter{\romannumeral2}} etc. \cite{Abgrall:2017syy,Alduino:2017pni,Albert:2017hjq,Agostini:2017hit}.
On the other hand, a $0\nu\beta\beta$ decay is possible through the diagram in Fig.~\ref{fig:Majoron}, when the composite neutrino state pair acquires a vacuum expectation value generating a Majorana mass, and then triggering the process.

\begin{figure}
\begin{centering}
\includegraphics[width=0.5\textwidth]{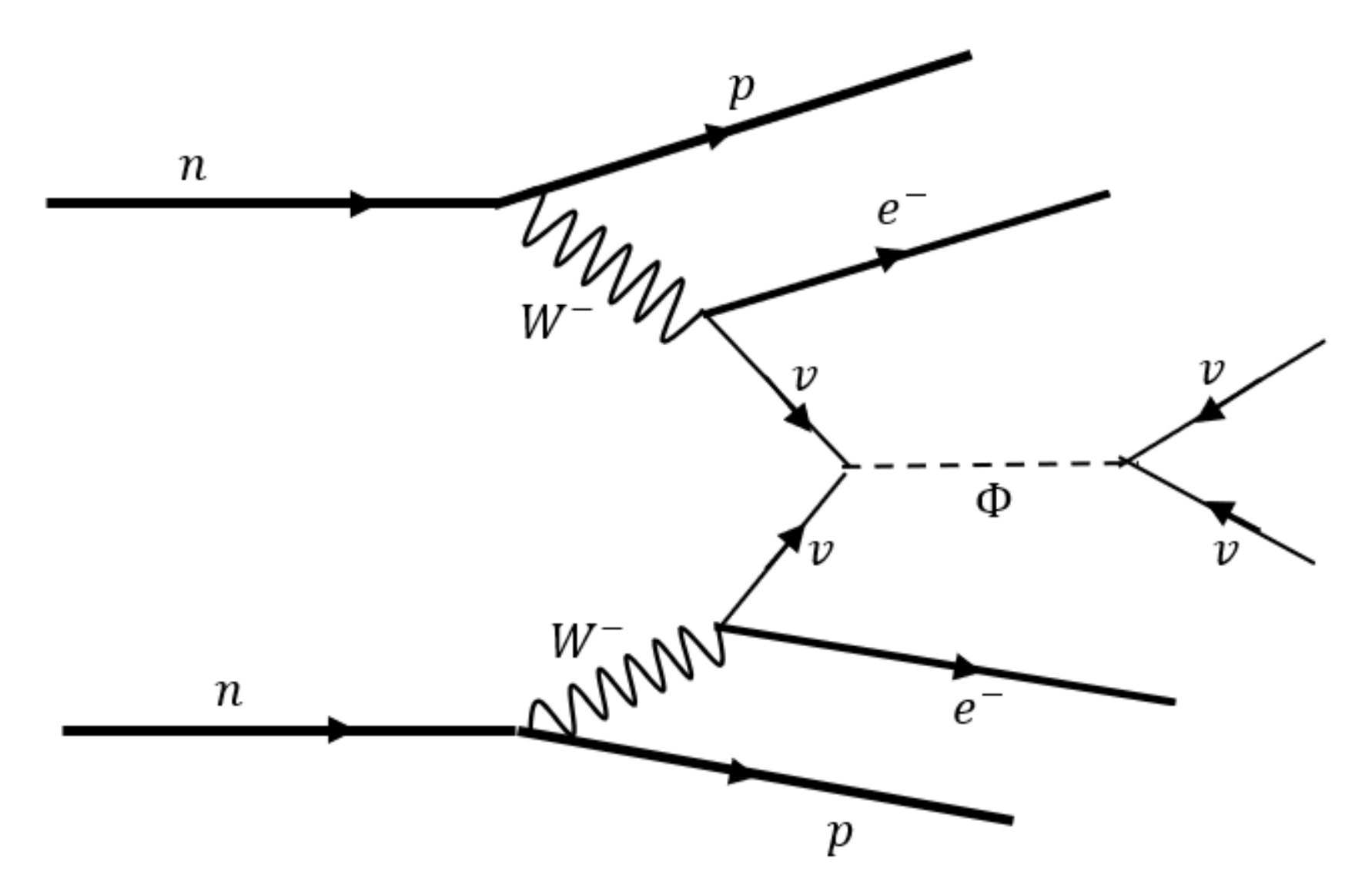}\includegraphics[width=0.5\textwidth]{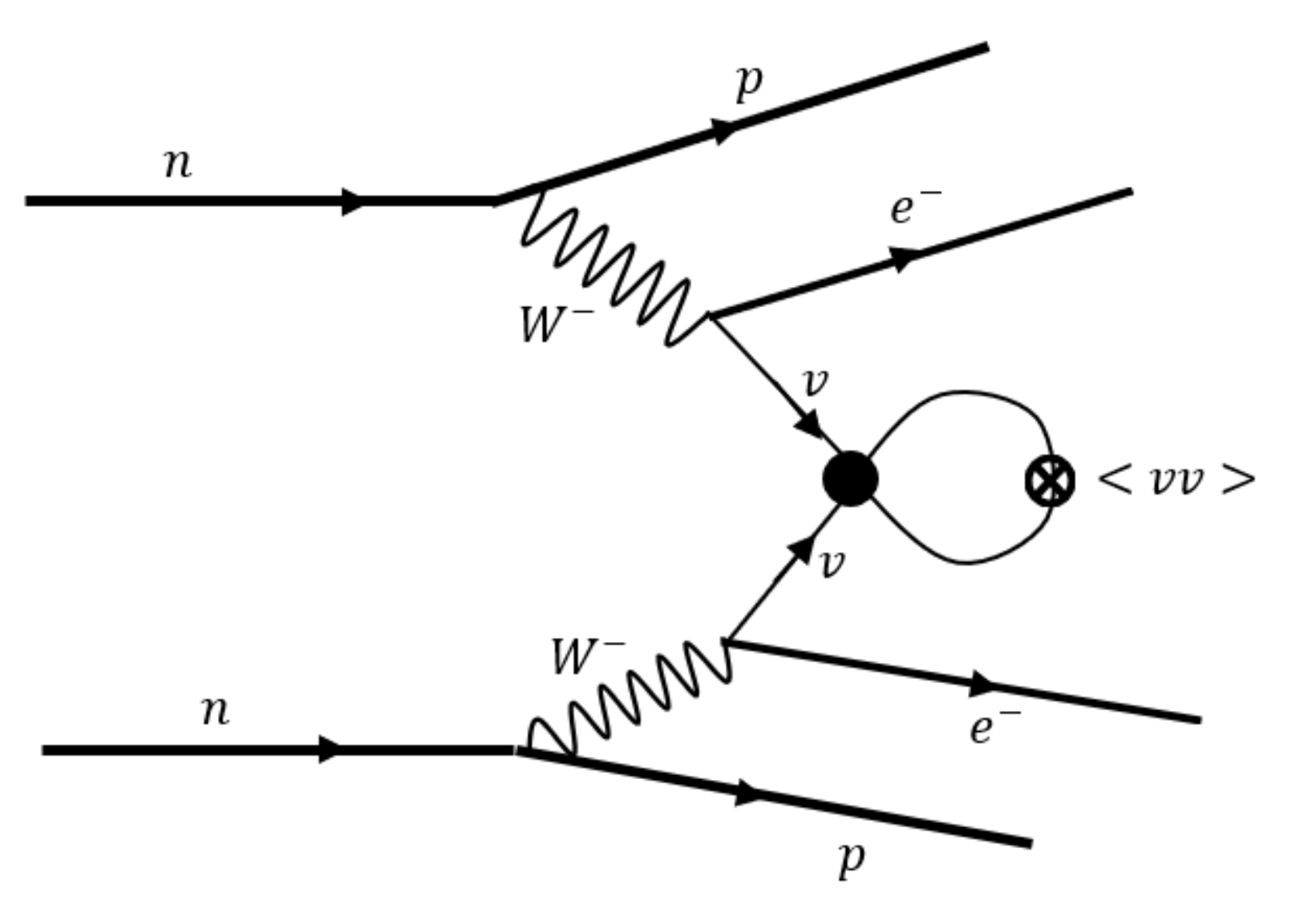}
\par\end{centering}
\begin{centering}
\caption{{\bf Left:} The emission of (anti)neutrinos through the composite Majoron channel is displayed. The emission of a bounded Majoron is kinematically impossible
and thus its detection from electron distribution alteration is forbidden. {\bf Right:} The $0\nu\beta\beta$ decay process through the neutrino pair condensation is displayed. }
\label{fig:Majoron}
\par\end{centering}
\end{figure}

\section{Discussion and Conclusions}

\label{sec:Conclusions-and-Remarks}

We have elaborated on a possible, common unifying explanation of Dark Energy, Dark Matter and the Neutrino mass origin. In particular, we have postulated the existence of a new dark fifth force interaction that we dub non-linear dark photon. The non-linear dark photon has a higher derivative electrodynamic Lagrangian, with particular interest to the Born-Infeld case. We have shown that non-linear higher derivative terms drive the new vector boson to a condensation with possibility to source the acceleration of the Universe. Then, we assume that only neutrinos are coupled to the new dark interaction. We have shown that the neutrino mass can be generated as an effect of the neutrino interaction with the Dark Energy Born-Infeld condensate.

This scenario opens the possibility of having a new state of neutrino matter: if neutrinos are produced as non-relativistic and cold in the early Universe, around the meV energy scale, they can cross a phase transition forming a neutrino superfluid. The neutrino superfluid is composed by Cooper pairs of misaligned spin neutrinos providing for a Majorana mass for the neutrinos. Indeed,
at meV energy, neutrinos are very weekly coupled to the SM particles, while are strongly coupled to the Dark Energy Born-Infeld condensate. On the other hand, mixing flavour neutrino pairs can provide non-diagonal mass terms sourcing neutrinos' oscillations \footnote{Neutrino oscillation has been intensively studied over the past two decades, within a series of investigations particularly focusing on the concomitant role of gravity and extended theories of gravity --- see e.g. Refs.~\citep{Capozziello:1999ww, Capozziello:1999qm, Capozziello:2000ga, Capozziello:2010yz, CaLa20202}.}. In principle, couplings of neutrinos to the Born-Infeld condensate are related to neutrino mass hierarchy. Differently than within several other see-saw models, a neutrino inverse hierarchy is naturally possible here.This is a topic that may result in great interest for the next neutrino experiments, including {\bf JUNO} \cite{An:2015jdp,Li:2014qca,Djurcic:2015vqa}.

Our work opens the pathway to several novel phenomenological possibilities to search for Dark Matter and test dynamical Dark Energy scenarios. First of all, if neutrino superfluid is responsible for Dark Matter, then it is possible to consider also the formation of Boson superfluid neutrino star. Consequently, the possible merging of neutrino Boson Stars may be observed through Gravitational Waves experiments. A similar reasoning applied to possible correlated signal of relativistic neutrino emissions generated by the high energy merging. 

We did not focus on identifying the precise detection range of 
of these phenomena, but we think this may deserve a future analysis beyond the purposes of this paper. Furthermore, several exotic scenarios may emerge from considering multi-operators induced by neutrino and Dark Energy, as for instance  $G_{F}'^{n}(\bar{\nu}\gamma_{\mu}\gamma_{5}\nu)^{2n}$
(spinor contracted combination). Within this latter case, one may have multi-neutrino pair condensate, pretty much reminding the already discussed possibility of Standard Model tetra-quarks, penta-quarks and so on. An alternative scenario may be that multi-neutrinos form nuggets of more compact $N>\!\!>1$ number with a binding energy scaling almost as $E_{Binding}\sim N M$ and a radius scaling as $R_{Binding}\sim1/(M\sqrt{N})$ \cite{Madsen:1998,Afshordi:2005ym,Gogoi:2020qif} .
For a large number of particles, neutrino nuggets can be compact and dark objects. This latter possibility opens the pathway to new exciting cosmological scenarios, which nonetheless are sill beyond our full comprehension.

\vspace{5mm}

\noindent \textbf{Acknowledgements.} A.A. work is supported by the Talent Scientific Research Program of College of Physics, Sichuan University, Grant No.1082204112427 $\&$ the Fostering Program in Disciplines Possessing Novel Features for Natural Science of Sichuan University, Grant No.2020SCUNL209 $\&$ 1000 Talent program of Sichuan province 2021. S.C. acknowledges the support of Istituto Nazionale di Fisica Nucleare, sez. di Napoli, iniziative specifiche QGSKY and MOONLIGHT2.
A.M. wishes to acknowledge support by the NSFC, through the grant No. 11875113, the Shanghai Municipality, through the grant No. KBH1512299, and by Fudan University, through the grant No. JJH1512105. Q.Y. Gan work is supported by the scholarship from China Scholarship Council (CSC) under the Grant CSC No. 202106240085.

\end{document}